\documentclass[iop]{emulateapj}

\usepackage{amsmath}
\usepackage{natbib}

\slugcomment{}

\newcommand{\be}{\begin{equation}}
\newcommand{\ee}{\end{equation}}

\shorttitle{Non-X-shaped Miras bulge}
\shortauthors{L\'opez-Corredoira}

\begin{document}

\title{Absence of an X-shaped Structure in the Milky Way Bulge Using Mira Variable Stars}
\author{Mart\'\i n  L\'opez-Corredoira\altaffilmark{1,2}}
\altaffiltext{1}{Instituto de Astrofisica de Canarias, E-38205 La Laguna, Tenerife, Spain; martinlc@iac.es}
\altaffiltext{2}{Departamento de Astrofisica, Universidad de La Laguna, E-38206 La Laguna, Tenerife, Spain}

\begin{abstract}
The stellar density distribution of the bulge is analyzed through one
of its tracers. We use oxygen-rich Miras variables from the Catchpole et al. (2016) survey 
and OGLE-III survey as standard candles. The average age of these stars is around 9 Gyr.
The population traced by Mira variables matches a boxy bulge prediction,
not an X-shaped one, because only one peak is observed in the density along the analyzed lines of sight, whereas the prediction of an X-shape gives two clear peaks.
\end{abstract}


\section{Introduction}


The nature of the Galactic bulge morphology has been discussed using different interpretations over the
last decades. Its non-axisymmetry was quickly recognized with the arrival of the
first near-infrared surveys (e.g., Weiland et al. 1994; L\'opez-Corredoira et al. 1997;
Babusiaux \& Gilmore 2005), but its shape has not been free from debate.
A peanut shape was evident in the projection of the bulge, especially in infrared surveys covering the full sky, like COBE-DIRBE, and 
extinction was considered to be responsible for the presence of the pronounced
``peanut shape'' waist located near $\ell=1^\circ $, $b=5^\circ $ in the
uncorrected DIRBE maps of the bulge at 1.25 and 2.2$\mu $m
(Weiland et al. 1994). From another perspective, astronomers have seen in this peanut shape some imprint of a boxy bulge  (e.g., Kent et al. 1991; Dwek et al. 1995; L\'opez-Corredoira et al. 2005) as observed in other galaxies (e.g., Shaw 1987) or predicted by some theories that argue that the boxiness of the bulge was a composite effect expected to appear when one considers stable orbits belonging to several families of periodic orbits (e.g., Combes \& Sanders 1981; Stiavelli et al. 1991; Merrifield 1996; Patsis et al. 2003).

Using similar images of the Milky Way in the infrared, some observers (Nataf et al. 2015; Ness \& Lang 2016) now see an X-shaped bulge behind the projected peanut shape, whereas other authors have seen elliptical or boxy bulges.
These, however, are not direct interpretations of the image but depend on the image processing with the subtraction of some particular disk model, or these may be artifacts from subtracting the bulge as an ellipsoid instead of as a boxy bulge (Lee \& Jang 2016). This follows
more recent theories of bulge formation (e.g., Patsis et al. 2002, Quillen et al. 2014, Li \& Shen 2015).
One may wonder why astronomers have not seen an X-shaped bulge in the 1990s in the COBE-DIRBE images.  
The structure along the line of sight has also been claimed to show an X-shapeusing metal rich red clumps as standard candles (Nataf et al. 2010, 2015; McWilliam \& Zoccali 2010; Saito et al. 2011; Wegg \& Gerhard 2013), but some doubts have been cast on whether the second peak along the line
of sight is a real density structure or an artifact in the luminosity
function of red clumps (Rattenbury et al. 2007; Lee et al. 2015; Joo et al. 2016; 
Lee \& Jang 2016; L\'opez-Corredoira 2016). There are also signs of a non-X-shape bulge in other populations:
very old and metal poor stars like RR-Lyrae in the bulge  are not 
X-shaped according to some authors (D\'ek\'any et al. 2013; Pietrukowicz et al. 2015), 
and young ($\lesssim 5$ Gyr) populations like F0-F5V stars are not X-shaped according to different interpretations (L\'opez-Corredoira 2016), interpretations that might also have
doubt cast upon. For instance, the argument for non-X-shape by L\'opez-Corredoira (2016) might also raise doubts based on the possible dispersion of absolute magnitudes, which might be much larger than assumed and thus may convolve the two peaks along the line of sight into one. If we assume that all of these results are correct, we would have the surprising result of a Galaxy with three bulges, which raises suspicions something is wrong with some of the previous analyses.

Here we want to throw further light on the question by analyzing a new population, Mira variables, which includes ages of both young and old stars.

\section{Distribution of Mira Variables}

Catchpole et al. (2016) report periods and JHKL observations for 643 oxygen-rich Mira variables found in two outer bulge fields centered at $b=-7^\circ $ and $\ell=\pm 8^\circ $,
each approximately 25 square degrees. 
They also use the 6528 Mira stars from the OGLE-III Catalog of Variable Stars
(Soszy\'nski et al. 2013) in regions with lower Galactic latitudes, providing
periods and photometry in V and I, which are dominated by oxygen-rich sources (Matsunaga et al. 2005). Here we use both sets of data and we cross-correlate OGLE data with 2MASS (Skrutskie et al. 2006) within 1 arcsec to also obtain the $K$-magnitude (in a single epoch), finding 6363 common stars.

Neither Catchpole et al. (2016) nor Soszy\'nski et al. (2013) have carried out a proper analysis to test the X-shape hypothesis, so these data can be further explored with that purpose.
Fig. 12 of Catchpole et al. (2016), with the intermediate-age population, is misleading since the clumpiness of the distribution is mostly due to noise: there are not enough stars to produce a map with bins of 0.15x0.15 kpc$^2$ (630 stars in $\sim 4$ kpc$^2$ gives an average of 3.5 star/bin, which is insufficient for performing a statistical analysis), and there are some incomplete counts because the entire selected region 
-6$^\circ<\ell<10^\circ$, $-4.5^\circ<b<-2.5^\circ $ was not totally covered. In any case, Catchpole et al. (2016) have not tested the X-shaped bulge hypothesis in this intermediate-age population. Instead, they just plotted Fig. 12 in this range of absolute magnitudes with the projected counts and put their results together with those of McWilliam \& Zoccali (2010), which correspond to $b=-8^\circ $, and observed that some stars are where they should be. The rough comparison between Fig. 12 and McWilliam \& Zoccali (2010) is invalid because they correspond to different latitudes.

Using the period--luminosity relationship (Whitelock et al. 2008; Catchpole et al. 
2016, Equation (2)) 
\begin{equation}
\label{mkp}
M_K=(-3.51\pm 0.20)\lambda-(7.25\pm 0.07)
,\end{equation}\[
\lambda=\log _{10}[P({\rm days})]-2.38
,\]
and the $K$-band extinction $A_K=0.11\,A_V$  (Fitzpatrick 1999). The
extinction  $A_V$ is calculated by Catchpole et al. (2016) for their data and 
by us using the Schlegel et al. (1998) maps for OGLE data; we can use this extinction to obtain the distance of each Mira star. 

\begin{figure}
\vspace{1cm}
\centering
\includegraphics[width=8cm]{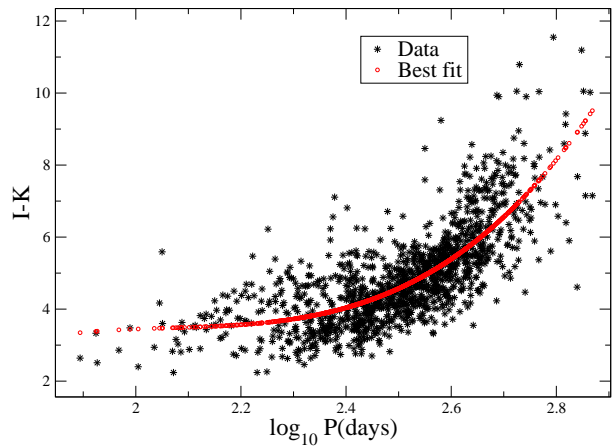}
\caption{Color $I$(OGLE)-$K$(2MASS) (corrected for extinction) 
in the sample of OGLE-III data with $|b|\ge 4^\circ $,
and the best fit of a cubic polynomial given by Equation (\ref{imk}).}
\label{Fig:imk}
\end{figure}

For the OGLE data, the $K$-magnitude is derived for a single epoch from 2MASS; it is not the average $K$-magnitude. This can generate a large dispersion in distances, due to the variability of the sources (Matsunaga et al. 2005). We wonder why Catchpole et al. (2016) did not make any attempt to correct
this dispersion of magnitudes, which introduces noise in the density maps. Here, in order to avoid this error in distance (although with average zero, which should be compensated for when large numbers of stars are used) due to the difference between the average magnitude of a Mira stars and the magnitude in a single epoch, we calibrate the average relationship between the color $I$(OGLE)-$K$(2MASS) (corrected for extinction) and the period 
in the OGLE sample of 1235 stars with $|b|\ge 4^\circ $ (to avoid high-extinction regions), and we use this relationship to calculate the ``average'' $K$-band magnitude in the OGLE stars used here.
This relationship is
\begin{equation}
I-K=3.96+3.69\lambda +10.33\lambda ^2+10.98\lambda ^3
\label{imk}
.\end{equation}
The dispersion of 2MASS single-epoch $K$-magnitudes with respect to this law gives an average rms=0.85 mag (see Fig. \ref{Fig:imk}), which
is on the order of magnitude of the expected value due to variability (Matsunaga et al. 2005), with higher dispersion toward higher periods. Given that we have used 1235 stars, the error of $I-K$ given in the above expression is 0.024 mag.
If we choose stars with $\log _{10}[P({\rm days})]<2.6$, the rms is reduced
to 0.62 mag. We will use this constraint in OGLE stars to avoid the highest deviations
from the color--period relationship.

Then, for the OGLE data,  rather than using the single-epoch 2MASS magnitude, we use the 
$I$-magnitude for each of the stars and this color
to calculate the average $K$-magnitude, i.e.
$m_K=m_I-(I-K)$, where $(I-K)$ is calculated from the period in Eq. (\ref{imk}).

\begin{figure*}
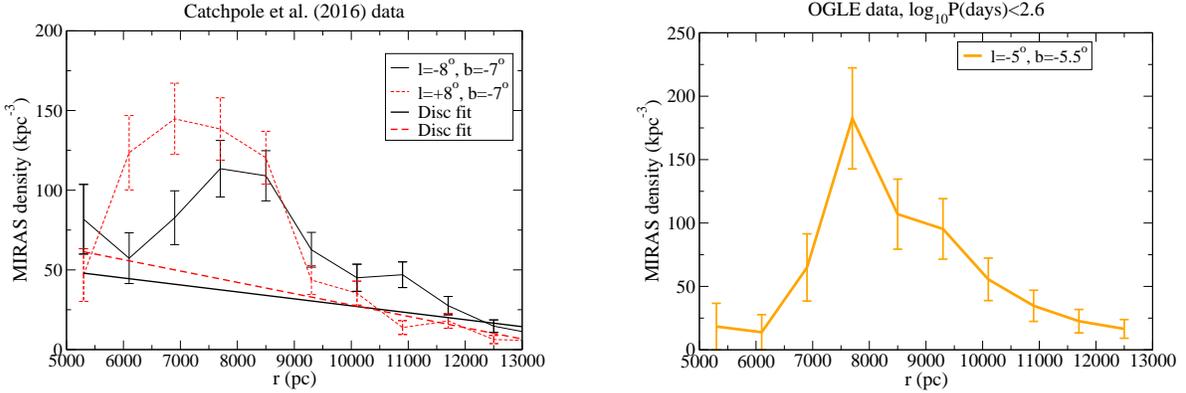

\vspace{1cm}
\centering
\includegraphics[width=7.2cm]{mirasdens.eps}
\hspace{1cm}
\includegraphics[width=7.2cm]{mirasogledens.eps}
\caption{Density of Mira variable stars along several lines of sight. 
In the left panel, an example of 
disk-component fitting at 4.5 kpc$<r<5$ kpc and 11 kpc$<r<14.5$ kpc is shown. 
Error bars are derived from
Poissonian errors.}
\label{Fig:mirasdens}
\end{figure*}

In Fig. \ref{Fig:mirasdens}, we see the distribution of Cathpole et al. (2016) along different lines of sight  and for some interesting regions of OGLE data.
For our analysis, since X-shape features are observable mainly at $4^\circ \lesssim |b| \lesssim 10^\circ $ (Nataf et al. 2010, Wegg \& Gerhard 2013), we study the region 
with the lowest negative longitude and the highest abs(latitude): 
$-7^\circ <\ell <-3^\circ$, $-6.5^\circ < b < -4.5^\circ $.
The density and star counts are related through 
$\rho [r(m_K)]=\frac{5}{\ln 10}\frac{1}{\omega \,r(m_K)^3}A(m_K)$, where
$r(m_K)=10^{[m_K-M_K+5]/5}$, $m_K$ is the extinction-corrected $K$-band magnitude and the differential star counts are $A(m_K)\equiv \frac{dN(m_K)}{dm_K}$.
There may be some incompleteness of sources due to the patchiness of extinction, especially in the regions with lower $|b|$, but this does not affect the shape of
the density along the line of sight, although its amplitude is affected.
In the same plot, we do a linear fit in the regions with 4.5 kpc$<r<5$ kpc and 11 kpc$<r<14.5$ kpc, which represents approximately the disk counts along those lines of sight, and the excess over this linear fit is the bulge density. 
The analysis of the disk counts is not the matter here, but note that this may be
composed of both thick and thin disks, and the inner flare of the disk (L\'opez-Corredoira et al. 2004) leads to a sizeable amount of disk stars even 
at $\sim 1$ kpc from the plane. 
For the bulge density, we can observe that there is only one peak in all lines of sight.

\begin{figure*}
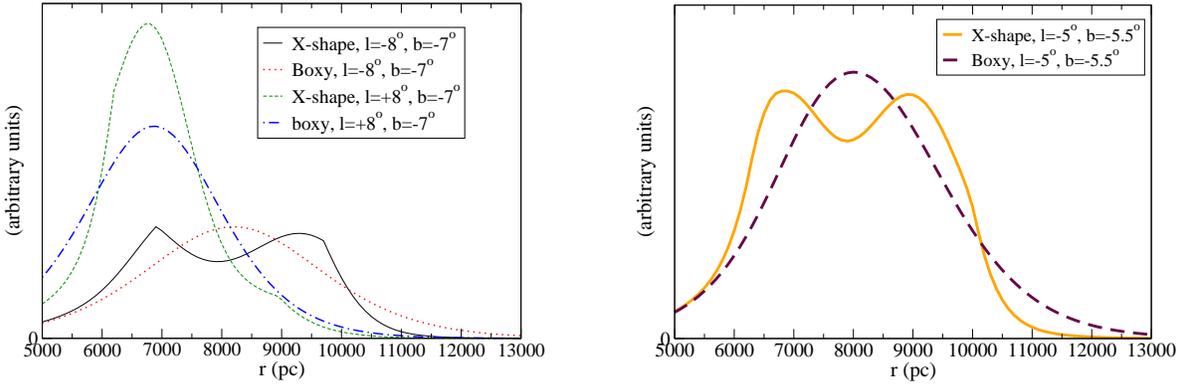

\vspace{1cm}
\centering
\includegraphics[width=7.2cm]{pred_miras.eps}
\hspace{1cm}
\includegraphics[width=7.2cm]{pred_mirasogle.eps}
\caption{Predictions for an X-shaped or a boxy bulge of the
density along the lines of sight of Fig. \ref{Fig:mirasdens}.}
\label{Fig:pred_miras}
\end{figure*}

A prediction of the bulge densities of two models, X-shaped and boxy, 
is given in Fig. \ref{Fig:pred_miras}. They were derived from the analytical expressions
given in L\'opez-Corredoira (2016, \S 4.1) for the boxy and X-shaped bulges observed 
by L\'opez-Corredoira et al. (2005) and Wegg \& Gerhard (2013) respectively:

\begin{equation}  
\rho _{\rm Boxy}(x,y,z)=\rho _0\,\exp\left(-\frac{\left(x^4+\left(\frac{y}{0.5}\right)^4+\left(\frac{z}{0.4}\right)^4\right)^{1/4}}{740\,{\rm pc}}\right)
\end{equation}

\[
\rho _{{\rm X-shape}}(x,y,z)=\rho _0\, \exp\left(-\frac{s_1}{700\,{\rm pc}}\right)\, \exp \left(-\frac{|z|}{322\,{\rm pc}}\right)
\]\begin{equation}
\label{densx}
\times \left[1+3\,\exp\left(-\left(\frac{s_2}{1000\,{\rm pc}}\right)^2\right)
+3\,\exp\left(-\left(\frac{s_3}{1000\,{\rm pc}}\right)^2\right) \right]
,\end{equation}\[
s_1={\rm Max}\left[2100\,{\rm pc},\sqrt{x^2+\left(\frac{y}{0.7}\right)^2}\right]
,\]\[
s_2=\sqrt{(x-1.5z)^2+y^2}
\]\[
s_3=\sqrt{(x+1.5z)^2+y^2}
.\]

For the positive longitude $\ell =+8^\circ $ there is not much distinction in the functional shape of the density between the both models. However,
for the negative longitude $\ell =-8^\circ $ line of sight and
for the three selected OGLE regions, the two predictions are very different,
with a double peak for the X-shaped bulge and a single peak for the boxy bulge. 

\begin{figure*}
\vspace{1cm}
\centering
\includegraphics[width=14cm]{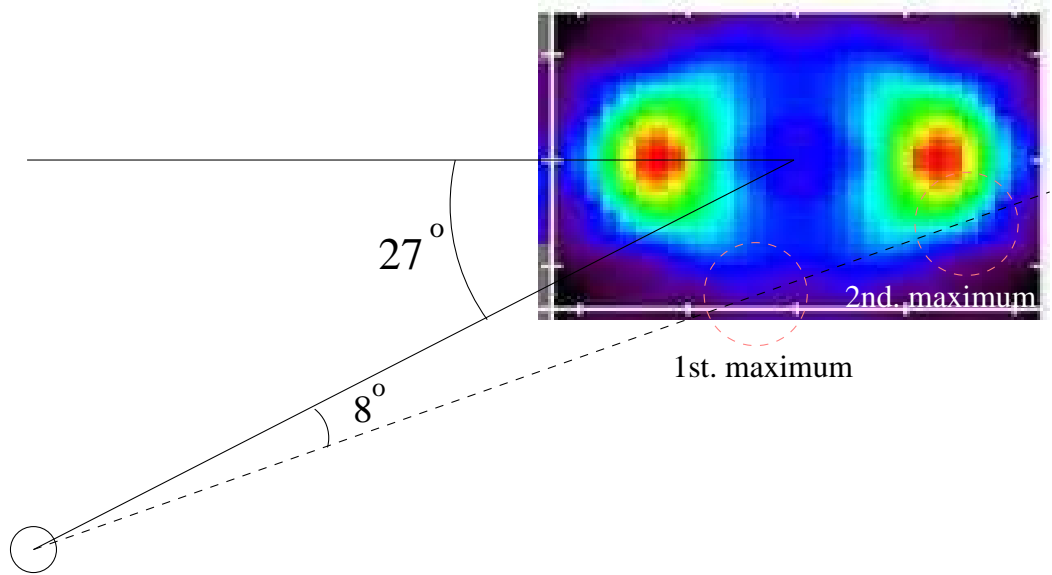}
\caption{Symmetrized density of the X-shaped model at $|z|=1012.5$ pc from Wegg \& Gerhard (2013), derived from red clumps, assuming a Galactocentric distance of 8 kpc and a bar angle
of 27$^\circ $. The dashed line stands for $\ell =-8^\circ $. Note how that this line crosses two maxima.}
\label{Fig:wegg}
\end{figure*}

Fig. \ref{Fig:wegg} shows a plot that helps us to understand the origin of this
double peak along $\ell =-8^\circ $, $b=-7^\circ $. This line of sight crosses two maxima in a constant $z$ slice, first in the left-hand bump, at around $r=7$ kpc, followed by a constant or slight decrease of the density in the hollow created by the peanut shape (the line remains in the region between the blue and violet colors), and when we take into account the increase of $|z|$ at higher values of $r$, this becomes a decrease of density. Then there is a second big peak at around 9.5 kpc associated with the right-hand bump; this second peak is indeed not as high in amplitude in Fig. \ref{Fig:pred_miras} because we must add the effect of the increase of $|z|$ with distance, which makes
the amplitude of the second peak similar to the first peak.
One may wonder why Fig. 6 of Wegg \& Gerhard (2013) and Fig. 3 of Saito et al. (2011) 
do not explictly show the double peaks. Indeed, these images show double bumps, but the lowest contours do not reach the longitude $-8^\circ $. It is a question of constrast, as it is not enough in these figures to see the lowest contours. However, in Fig. 3/top-right panel of McWilliam \& Zoccali (2010) for ($\ell=-6^\circ $, $b=-8^\circ $), a double peak is clearly observed; see also their Fig. 6, which shows the effect of double peaks between $\ell=-9^\circ $ and $\ell=+3^\circ $.
For lower values of $|\ell |$, it is even more evident that the X-shape should provide double
peaks at negative longitudes.
In any case, if there is no double peak for the red clumps, it 
would remove the question regarding the X-shape topic at least for this region, although not for the other regions.

Clearly, the boxy
bulge resembles the results of Fig. \ref{Fig:mirasdens}, whereas the X-shaped bulge does not. 
For the OGLE data, the covered areas are lower, and that is the reason it produces a
noisier density.
A $\chi^2$ analysis in the comparison of the models with the data in the range 5 kpc$<r<11$ kpc (once subtracted the disk) gives
\begin{enumerate}
\item for the line of sight $\ell=-8^\circ $, $b=-7^\circ $: 
$\chi ^2_{\rm reduced}\equiv \frac{\chi ^2}{N-1}$ ($N=8$ values of $r$)
 equal to 2.81 and 1.09,
respectively, for the X-shaped and boxy bulges, corresponding to probabilities of
$6.4\times 10^{-3}$ and 0.37, i.e. the X-bulge is excluded within 99.36\% C.L. ($2.7\sigma $) whereas the boxy bulge matches the data; and

\item for the line of sight $\ell=-5^\circ $, $b=-5.5^\circ $:
$\chi ^2_{\rm reduced}$
 equal to 2.30 and 1.40,
respectively, for the X-shaped and boxy bulges, corresponding to probabilities of
$0.024$ and 0.20, i.e. the X-bulge is excluded within 97.6\% C.L. ($2.3\sigma $) 
and the boxy bulge fit is not good but it is only excluded within
80\% C.L. (1.3$\sigma $), which is much likely than the X-shaped bulge.

\end{enumerate}

The combination of the two lines of sight gives,
respectively, for the X-shaped and boxy bulges a probability
of $1.1\times 10^{-3}$ and 0.23, i.e. the X-bulge is excluded within 99.89\% C.L. ($3.3\sigma $) whereas the boxy bulge is only excluded within 77\% C.L. ($1.2\sigma $), which
can be considered as a likely fit.

The average relative error of distance of these Mira stars using Eq. (\ref{mkp}) is 3.6\% (Catchpole et al. 2016), which is too low to produce a smoothing of the two peaks into one.
The error in Eq. (\ref{imk}) is also low, as mentioned, around 0.024 mag. The
error in the extinction is around 10\% (Schlegel et al. 1998), with the average extinction
$A_K\approx 0.2$ mag for OGLE regions with $|b|\ge 4^\circ $. This produces another error
of $\Delta A_K\sim 0.02$ mag and $\Delta A_I\sim 0.1$ mag, which is again too low to produce
any confusion of the two peaks. The effect of metallicity dispersion is also negligible: first because the
variations of the metallicity in the bulge along a line of sight are very low (L\'opez-Corredoira
2016), and second because even large variations of metallicity would produce variations of the magnitude of less than 0.1 mag (Matsunaga 2012), which would mean an 
error lower than $0.1/\sqrt{N}$ mag in each bin with $N$ the number of stars in that bin.
This makes this standard candle much more precise in the distance determination than the F0-F5V stars used by L\'opez-Corredoira (2016), but the result
is still the same: a single peak along the line of sight.

\begin{figure}
\vspace{1cm}
\centering
\includegraphics[width=8.5cm]{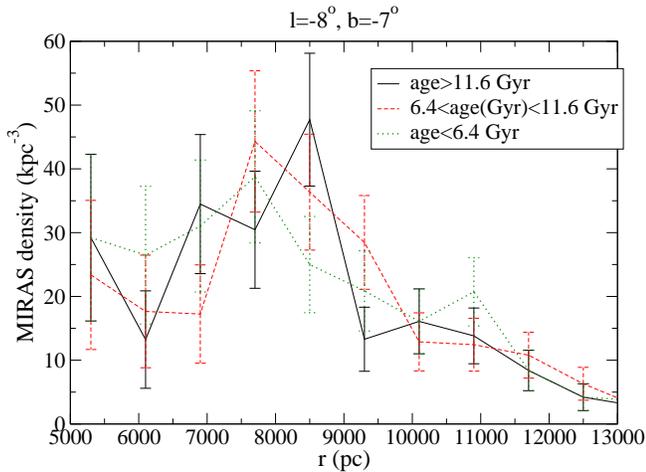}
\caption{Density of Miras variable stars along the line of sight
$\ell =-8^\circ $, $b=-7^\circ $ for different ranges of ages.}
\label{Fig:mirasdensage}
\end{figure}

The $\log _{10}P({\rm days})$ is related to the age of the star. Roughly,
the ages of 1.0, 1.8, 3.0, 5.0, 7.0 Gyr correspond to 
$\log _{10}P({\rm days})=2.83, 2.70, 2.65, 2.60, 2.50$ (Feast 2009; Catchpole et al. 2016) so an approximate relationship can be given by a second-order
polynomial fit of these numbers: 
\begin{equation}
\label{age}
{\rm age(Gyr)}\sim 12-44\lambda+43\lambda ^2
.\end{equation} 
The average $\log _{10}P$ of the Mira variables in Fig. \ref{Fig:mirasdens} 
is 2.46 at the ($\ell=-8^\circ $, $b=-7^\circ $) line of sight and 2.45 for the 
combination of the three OGLE lines of sight, 
which represents, according to Eq. (\ref{age}), average ages of 8.8 and 9.1 Gyr
respectively.
In Fig. \ref{Fig:mirasdensage}, we represent the density 
along ($\ell=-8^\circ $, $b=-7^\circ $)
for three bins with approximately the same number of stars and ranges of
$\log _{10}P$ ranges of $<2.39$, 2.39-2.53 and
$>2.53$, which correspond, using Equation (\ref{age}),
to ages of $\gtrsim 11.6$ Gyr, 6.4-11.6 Gyr, and $\lesssim 6.4$ Gyr.
The C.L.s at which X-shapes are excluded in a $\chi ^2$ analysis are
2.1$\sigma $, 1.2$\sigma $, and 1.1$\sigma $, respectively, for these ages;
for the boxy bulge, it is excluded at $1.3\sigma $,
0.8$\sigma $, and 0.5$\sigma $, respectively. 
The only significant exclusion 
here is for the oldest Miras to be X-shaped. 
For the OGLE line of sight, the star counts are much
lower, so these data are even more inconclusive.
We would need a higher number of Mira stars
to be able to distinguish the shapes of the bulge at different ages. 

\section{Conclusions}

The bulge with an X-shaped morphology traced by the average population of Mira variables   is excluded at the 3.3$\sigma $ level, whereas a boxy bulge density distribution matches the data.
The significance is reduced to 2.3$\sigma $ if we exclude the region of $\ell=-8^\circ $, $b=-7^\circ $.
Apparently, there is no agreement regarding the structure of red clumps for this line of sight of $\ell=-8^\circ $, $b=-7^\circ $: some authors  obtained a double peak there (e.g, McWilliam \& Zoccali 2010), others do not see it there (e.g., Saito et al. 2011), others see it but relatively noisy (e.g., Nataf et al. 2015). The fiducial model of the X-shaped bulge used here (Equation (\ref{densx})) 
reproduces the Wegg \& Gerhard (2013) maps derived from red clumps. In any case, in our analysis using Mira stars, we have not seen double peaks along the line of sight,
not in $\ell=-8^\circ $, $b=-7^\circ $, nor in $\ell=-5^\circ $, $b=-5.5^\circ $.

This population has an average age of 9 Gyr. At present, using Mira variable stars we cannot distinguish the morphology of stars with different ages. 
This casts some doubt on the validity of red clump analyses claiming the existence of an X-shaped structure among stars with similar ages to those analyzed here, although an analysis of the metallicity of these stars is still pending. 
Bulge dynamics, formation, and evolution do not distinguish among the different types of stars with the same origin (same age and metallicity). Stars with the same origin should belong to the same structure. RR Lyrae are thought to be part of a different bulge compared to red clumps because RR Lyrae are older. Also, the F0-5V stars are younger than red clumps on average. But now we have a population of stars with ages similar to the red clumps with high metallicity, it is not expected that all stellar populations are part of a boxy bulge except for the red clumps, which are furthermore suspected to be contaminated by the second peak, unless we demonstrate a different origin for these red clumps.
Even if we cannot be 100\% sure, at the very least, doubts regarding the very existence of the X-shaped bulge should be established from the present work.

\begin{acknowledgements}
Thanks are given to the anonymous referee for the critical comments that
motivated new calculations that reinforced the robustness of the analysis to be carried out.
This work has been supported by the grant AYA2015-66506-P from the Spanish Ministry of Economy and Competitiveness (MINECO).
\end{acknowledgements}

\end{document}